\documentclass[a4paper,11pt]{article}
\pdfoutput=1 

\usepackage{jheppub} 

\usepackage[T1]{fontenc} 

\usepackage{amsfonts}
\usepackage{graphics}
\usepackage{epsfig}
\usepackage{url}
\usepackage{multirow}
\usepackage{feynmp}
\newcommand {\be}{\begin{equation}}
\newcommand {\ee}{\end{equation}}
\newcommand {\ba}{\begin{eqnarray}}
\newcommand {\ea}{\end{eqnarray}}

\newcommand {\tanb}{$\tan\beta~$}
\newcommand {\ra}{\rightarrow}

\title{\boldmath Single Top Events as a Source of Light Charged Higgs in the Fully Hadronic Final State at LHC}

\author[]{M. Hashemi}

\affiliation[]{Physics Department and Biruni Observatory,\\College of Sciences, Shiraz University, Shiraz 71454, Iran}

\emailAdd{hashemi$\_$mj$@$shirazu.ac.ir}

\abstract{
The single top production proceeds through different channels at LHC. The $t$-channel production which has a cross section at the level of one third of the $t\bar{t}$, can be regarded as a source of light charged Higgs boson through top quark decay to charged Higgs. Due to the low jet multiplicity and characteristic kinematic features of this process, a signal selection can be designed to extract charged Higgs signal from the background processes. In this paper, taking MSSM as the theoretical framework, The light charged Higgs signal from single top events is analyzed. The charged Higgs is assumed to decay to $\tau\nu$ and the hadronic decay of $\tau$ leptons is analyzed. Different charged Higgs mass hypotheses are analyzed and a scan over the parameter space ($m_{H^{\pm}},tan\beta$) is performed. It is shown that this process has the potential to exclude large areas of parameter space, while with enough data, a 5$\sigma$ discovery would also be possible. Finally 95$\%$ C.L. exclusion and 5$\sigma$ contours are provided for different integrated luminosities of LHC.}

\begin{document} 
\maketitle
\flushbottom

\section{Introduction}
In the minimal supersymmetric extension of standard model of particle physics, five Higgs bosons are predicted, two neutral CP-even bosons, $h^0,~H^0$, a neutral CP-odd boson, $A^0$, and two charged Higgs bosons, $H^{\pm}$. Neutral Higgs bosons may appear like their standard model partner. The lightest one, $h^0$ is usually taken to be SM-like, with decay properties coinciding those of SM Higgs boson. However, the other two neutral Higgs bosons differ from their SM partner, in terms of their decay channels and branching ratio of decays. Among the Higgs boson family members of MSSM, the charged Higgs boson plays an important role in the search for MSSM signals. While being charged, it provides a unique signature compared to neutral Higgs bosons of MSSM.\\
There have been  numerous searches for this particle during the last years at Tevatron and LEP. Limits are quoted in terms of the mass of this particle as a function of \tanb which is the ratio of vacuum expectation values of the two Higgs fields used to make the two Higgs doublets. The LEP II experiment, in a direct search, has excluded charged Higgs mass lighter than $89~ \textnormal{GeV}$ for all \tanb assumptions \cite{lepexclusion1}. The indirect limits are however stronger, excluding charged Higgs masses lighter than $125~ \textnormal{GeV}$ \cite{lepexclusion2}. \\
The Tevatron searches by the D0 \cite{d01,d02,d03,d04} and CDF Collaborations \cite{cdf1,cdf2,cdf3,cdf4} allow \tanb to be in the range $2 <$ tan$\beta < 30$ for $m(H^{\pm}) > 80 $ GeV while more $\tan \beta$ values are available for higher charged Higgs masses.\\
There are B-Physics constraints obtained for general two Higgs doublet models (2HDM). The CLEO data excludes a charged Higgs mass below 295 GeV at 95 $\%$ C.L. in 2HDM Type II with \tanb higher than 2 \cite{B1}. However, such constraints belong to general 2HDMs and in general, a translation of those constraints frmo B-Physisc studies to the case of MSSM may not be trivial. \\ 
The most recent results on the charged Higgs searches by ATLAS and CMS experiments at LHC \cite{atlasdirect,cmsdirect} indicate that a light charged Higgs with $m_{H^{\pm}}\sim 125 ~\textnormal{GeV}$ is excluded with \tanb $\succsim 10$ (ATLAS) and \tanb $\succsim 20$ (CMS). The ATLAS exclusion is stronger, however, as it is not yet confirmed by CMS analyses, we use CMS results (\tanb $\succsim 20$) for comparisons with our results. The starting point taken in this paper is $m_{H^{\pm}}\sim 125 ~\textnormal{GeV}$, i.e., the analysis scans charged Higgs masses above 125 GeV.\\
\section{Single Top Events as a Source of Charged Higgs Bosons}
\begin{figure}
\centering
\includegraphics[width=0.7\textwidth]{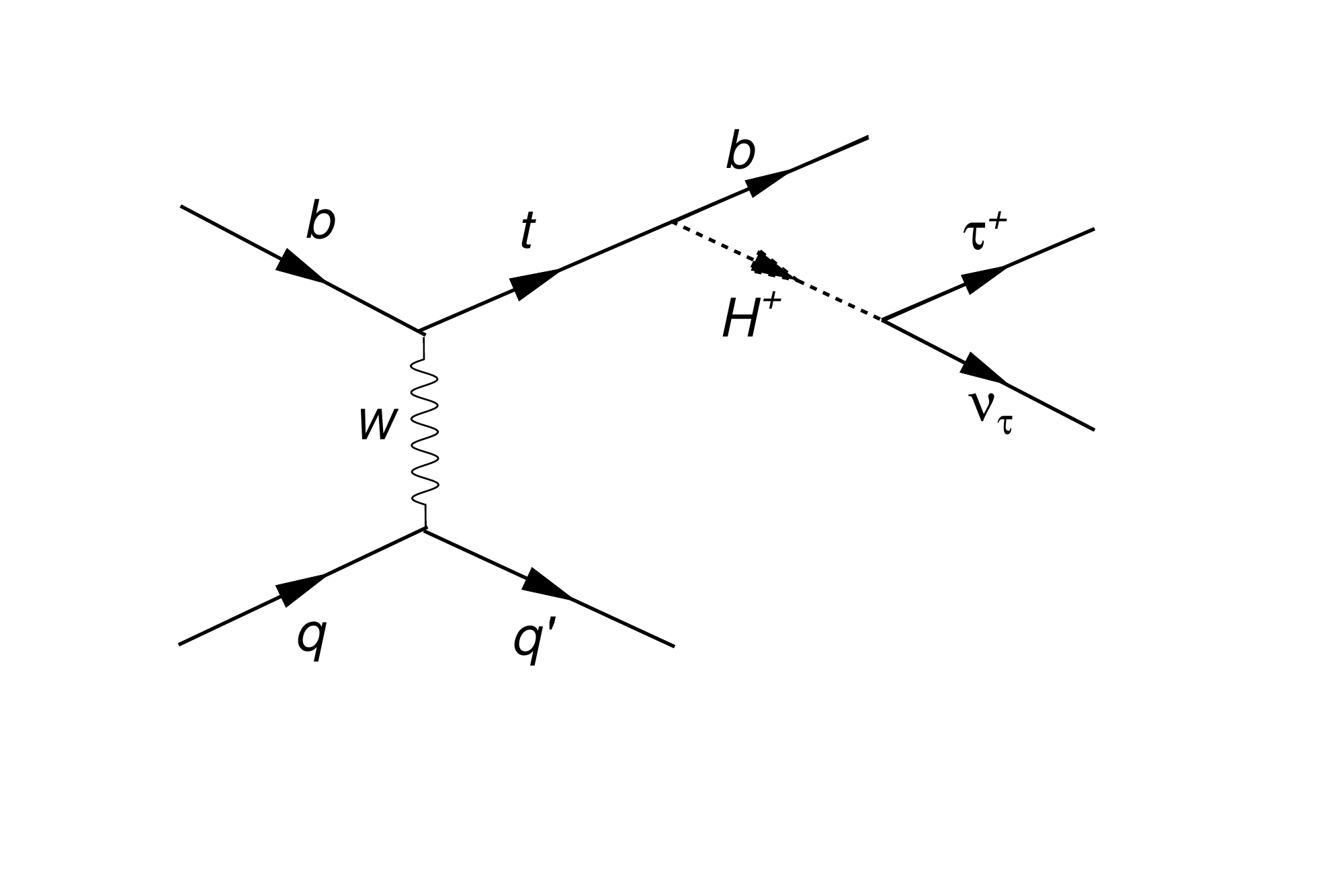}
\caption{The full signal production chain: the $t-$channel single top diagram as the signal process followed by the top quark decay to charged Higgs which in turn decays to $\tau$ lepton.\label{diagram}}
\end{figure}

The single top production process proceeds through three different channels, i.e., the $s$-channel single top, the $t$-channel single top and $tW$ channel. The top quark in $s$-($t$-)channel, is produced in association with a $b$ (light) quark. First observation of single top events was announced by D0 \cite{d0st} and \cite{cdfst} collaborations. The ATLAS and CMS collaborations have also reported observation and cross section measurement of the single top in \cite{atlasst,cmsst}. \\
This process is generally attractive for several reasons. It is an electroweak process which can be used to acquire information about the CKM matrix entry $V_{tb}$. Low multiplicity in the final state paritcle collection is a superiority of this process compared to the top pair events which are characterized by their large number of final state particles. This feature has two consequences. First, a better particle isolation and identification efficiency is expected in single top events compared to top pair events. Second, single top events are less dependent on the systematic uncertainties due to containing less number of physical objects in the event.\\
Due to the low cross section of single top events at Tevatron, a detailed study of its properties was made available when a large amount of data was finally collected. However, the cross sections of this process at LHC in three $s$, $t$ and $tW$ channels are as large as $10.5,247$ and $51~pb$ \cite{topquarkphysics}. Since the $t$-channel single top has a large cross section (almost one third of $t\bar{t}$ cross section), it can be used as the main single top channel for different studies, one of which could be search for light charged Higgs bosons through the subsequent decay $t\ra H^{+}b$. In addition to the aforementioned reasons for the study of single top events, results of this analysis can be considered as a confirmation of LHC results obtained from top pair events or can be used in a statistical combination, to make the final conclusions on exclusion/discovery contours stronger.\\
This channel has not yet been analyzed at LHC using real data in the search for charged Higgs bosons, however, as will be seen in this paper, it provides reasonable results compared to the already obtained results from top pair events even with a basic set of cuts. Phenomenological studies of single top events have already been appeared in the literature. In \cite{oldreview}, single top events have been analyzed in a search for the light charged Higgs decaying to $\tau$ leptons which in turn undergo leptonic decays. Updated results are reported in \cite{st1,st2} where the same final state has been analyzed. Results quoted in terms of exclusion contours in ($m_{H^{\pm}},tan\beta$) are promising and motivate full analysis of single top events as a potential source of charged Higgs bosons at the presence of detector effects using real data at LHC. Analyses of heavy charged Higgs have also been performed at Tevatron \cite{st3}, where, the charged Higgs boson is produced alone and decays to a top and a bottom quark. In this analysis, the attempt is to analyze the light charged Higgs signal originating from $t$-channel single top events (Fig. \ref{diagram}). 
The final state is, however, different from that of \cite{st1,st2}, i.e., a fully hadronic final state is assumed. Description of signal and background events comes in the next section. The theoretical framework is based on MSSM, $m_{h}$-max scenario with the following parameters: $M_{2}=200$ GeV, $M_{\tilde{g}}=800$ GeV, $\mu=200$ GeV and $M_{SUSY}=1$ TeV.
\section{Signal and Background Processes and their Cross Sections}
The signal process is taken as the $t$-channel single top followed by the decay of the top quark to charged Higgs ,i.e., $pp \ra tq \ra H^{\pm}bq \ra \tau \nu b q$. The hadronic decay of the $\tau$ lepton is analyzed. The $s$-channel process has a small cross section and can be neglected as a signal source. The main background processes are $W^{\pm}jj$, $W^{\pm}b\bar{b}$, $W^{\pm}c\bar{c}$ (which are called ``W+2jets'' hereafter) and $t\bar{t}$. The cross sections of these processes are listed in Tab. \ref{xsec}. For the signal and $t\bar{t}$ cross sections MCFM 6.1 \cite{mcfm} is used while ALPGEN \cite{alpgen} is used for W+2jets generation and cross section calculation with a kinematic preselection applied as in Eq. \ref{presel}.
\be
E_{T}^{\textnormal{jet}}>20~ \textnormal{GeV}~~,~~|\eta^{\textnormal{jet}}|<5.0
\label{presel}
\ee
The parton distribution function (PDF) used in the analysis is CTEQ 6.6 which is provided by LHAPDF 5.8.3 \cite{lhapdf}. The signal and $t\bar{t}$ event generation is done using PYTHIA 8.153 \cite{pythia}. The W+2jets events are produced using ALPGEN whose output in LHA format is passed to PYTHIA for parton showering and hadronization. Jet reconstruction is then performed with FASTJET 2.4.1 \cite{fastjet} using anti-kt algorithm \cite{antikt} and $E_{T}$ recombination scheme with a cone size of $\Delta R=0.5$. Here $\Delta R=\sqrt{(\Delta\eta)^2+(\Delta\phi)^2}$ with $\eta=-\ln\tan\theta/2$ and $\theta (\phi)$ are polar (azimuthal) angles with the beam pipe defined as the $z$-axis.    
\begin{table}[h]
\begin{center}
\begin{tabular}{|c|c|c|c|c|c|c|c|}
\hline
Process & $t$-channel single top & $s$-channel single top & $t\bar{t}$ & $Wjj$ & $Wb\bar{b}$ & $Wc\bar{c}$ \\
\hline
Cross section [$pb$]& 247 & 10.5 & 834 & 1670 & 3.5 & 3.3 \\
\hline
\end{tabular}
\end{center}
\caption{Cross sections of signal and background processes \label{xsec}}
\end{table}
\section{Event Selection and Analysis}
\begin{figure}
\begin{center}
\includegraphics[width=0.6\textwidth]{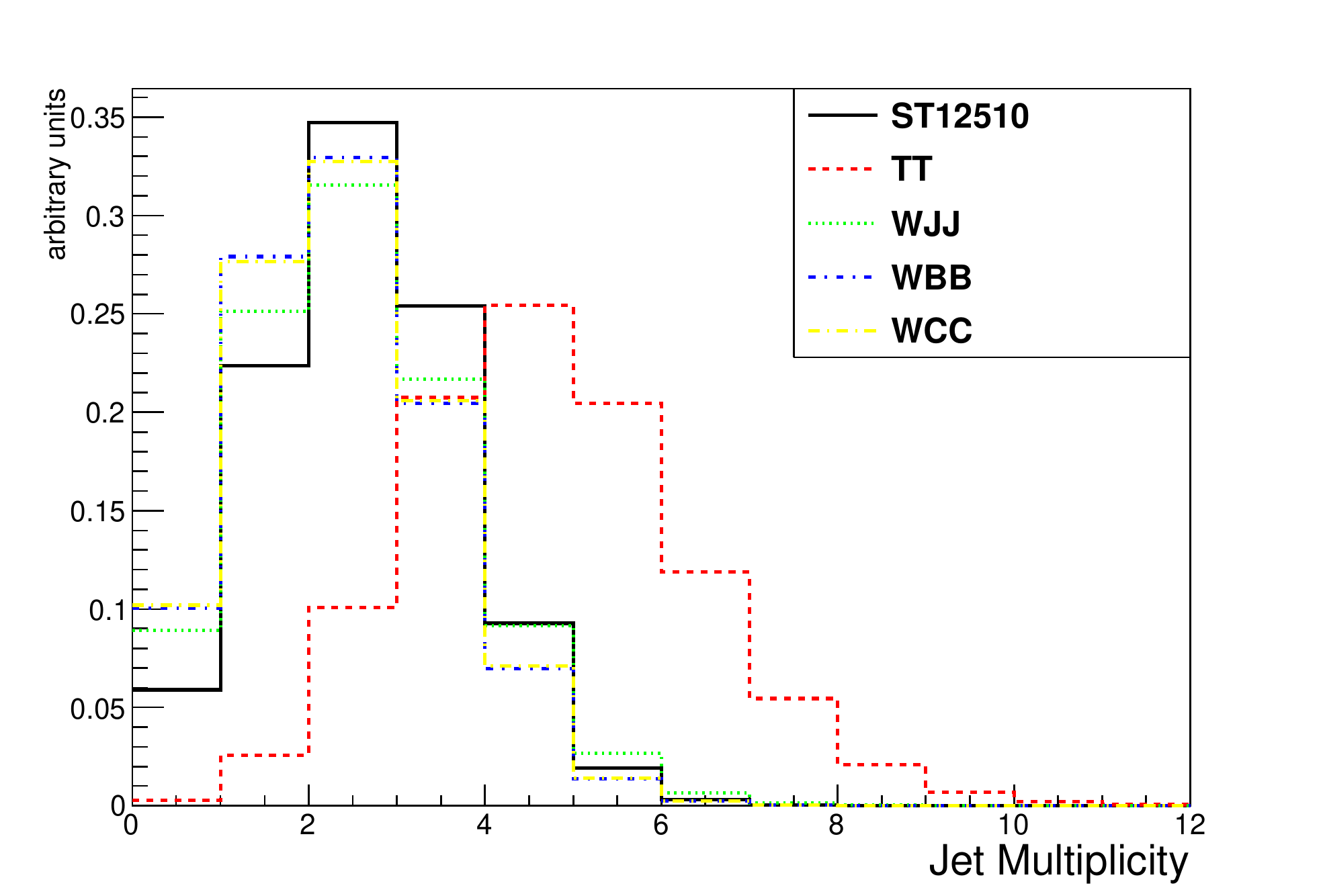}
\end{center}
\caption{Jet multiplicity distribution of signal and background events.}
\label{jm}
\end{figure}

\begin{figure}
\begin{center}
\includegraphics[width=0.6\textwidth]{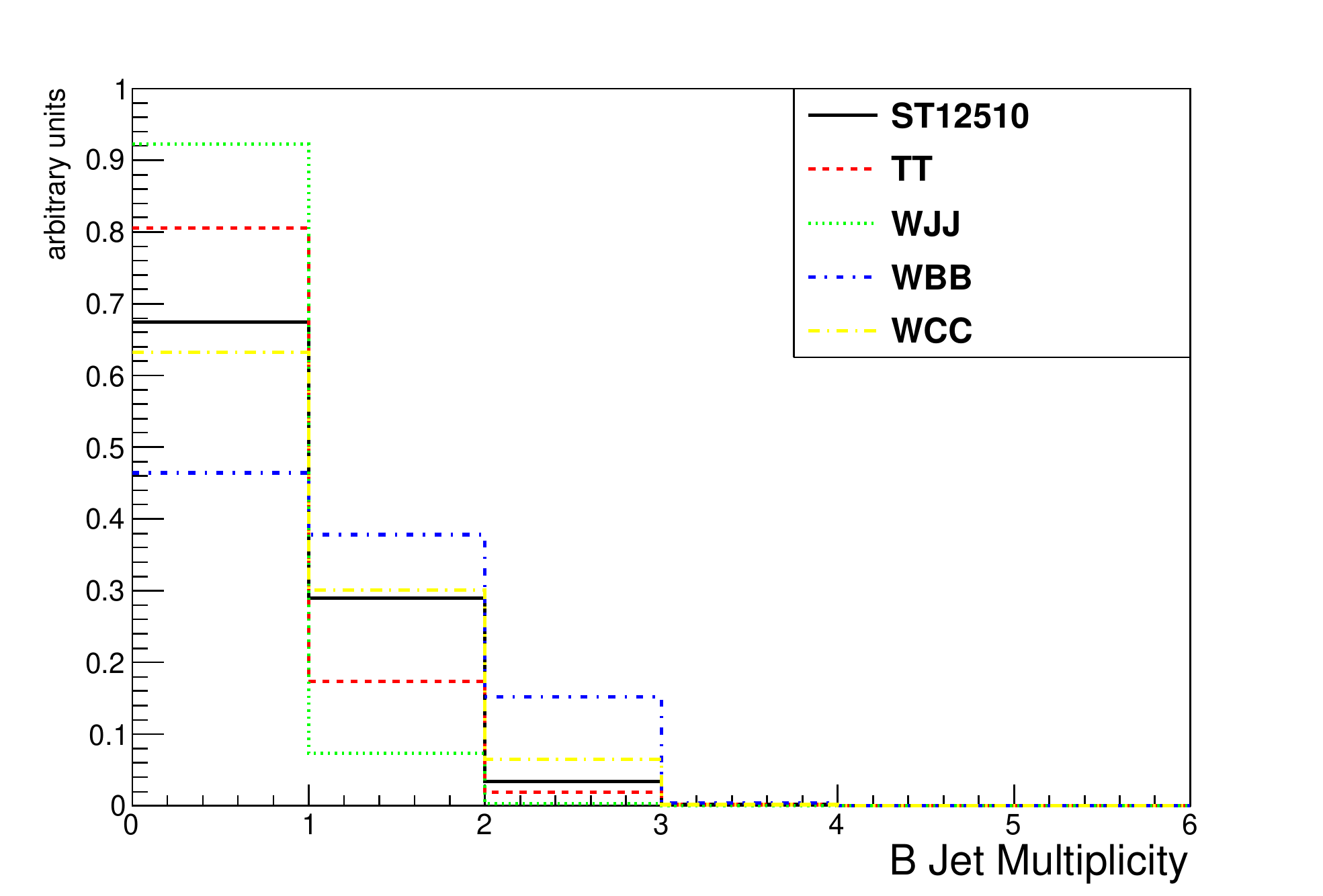}
\end{center}
\caption{The $b$-jet multiplicity distribution of signal and background events.}
\label{bjm}
\end{figure}

\begin{figure}
\begin{center}
\includegraphics[width=0.6\textwidth]{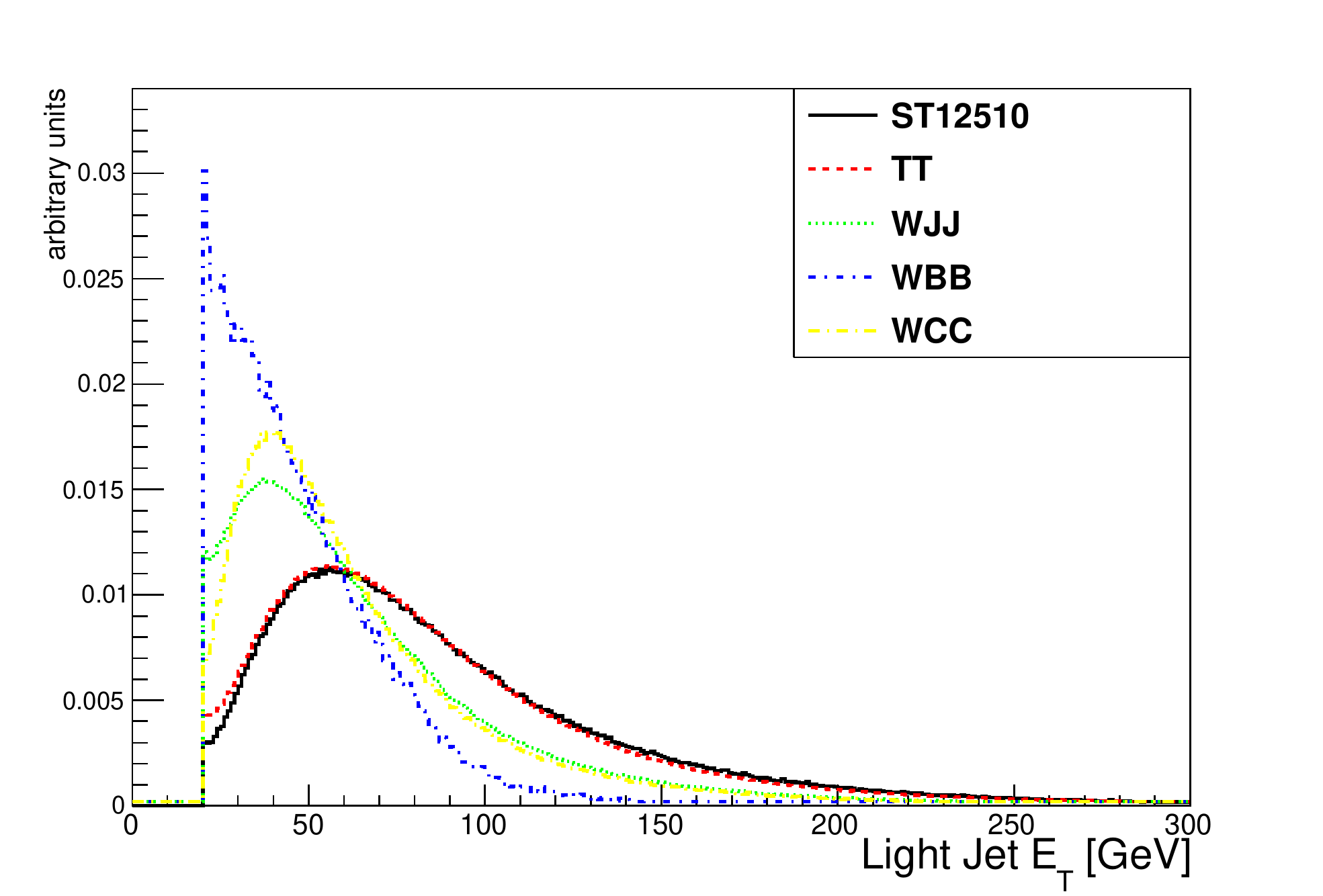}
\end{center}
\caption{The light jet multiplicity distribution of signal and background events.}
\label{ljet}
\end{figure}

\begin{figure}
\begin{center}
\includegraphics[width=0.6\textwidth]{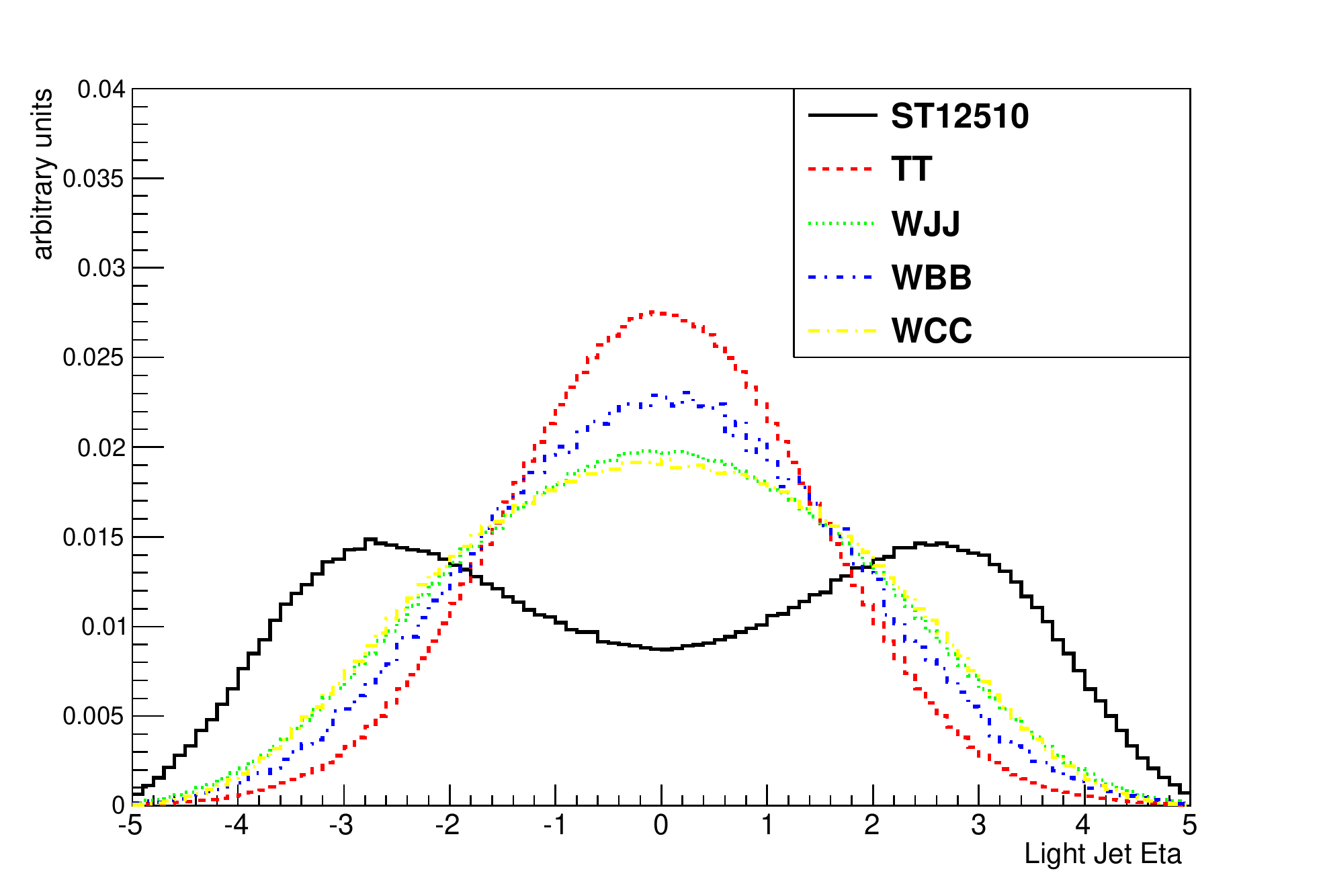}
\end{center}
\caption{The light jet multiplicity distribution of signal and background events.}
\label{ljeta}
\end{figure}

\begin{figure}
\begin{center}
\includegraphics[width=0.6\textwidth]{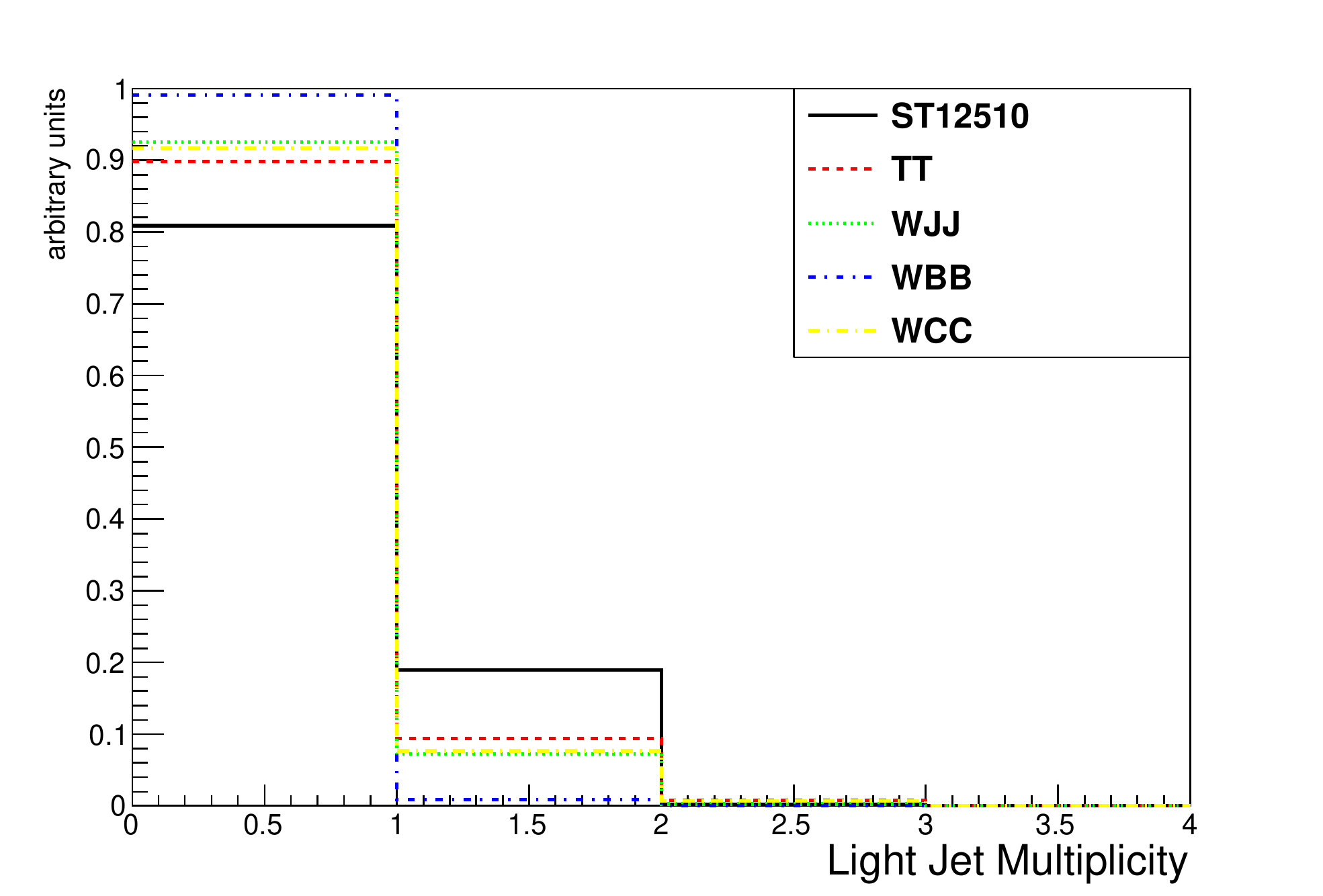}
\end{center}
\caption{The light jet multiplicity distribution of signal and background events.}
\label{ljetmul}
\end{figure}

\begin{figure}
\begin{center}
\includegraphics[width=0.6\textwidth]{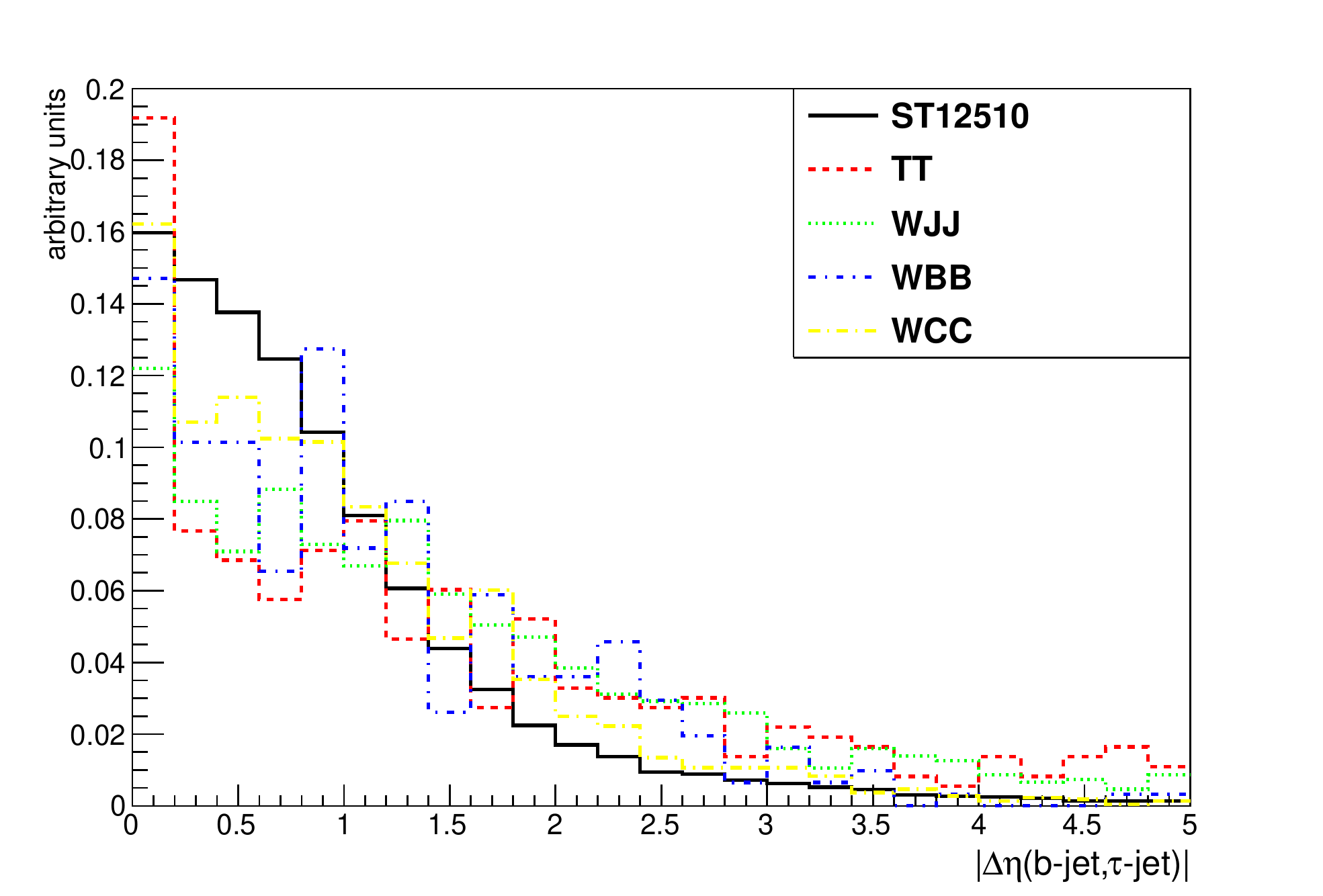}
\end{center}
\caption{$\Delta\eta$ between the $b-$jet and the $\tau-$jet in signal and background events.}
\label{detabt}
\end{figure}

\begin{figure}
\begin{center}
\includegraphics[width=0.6\textwidth]{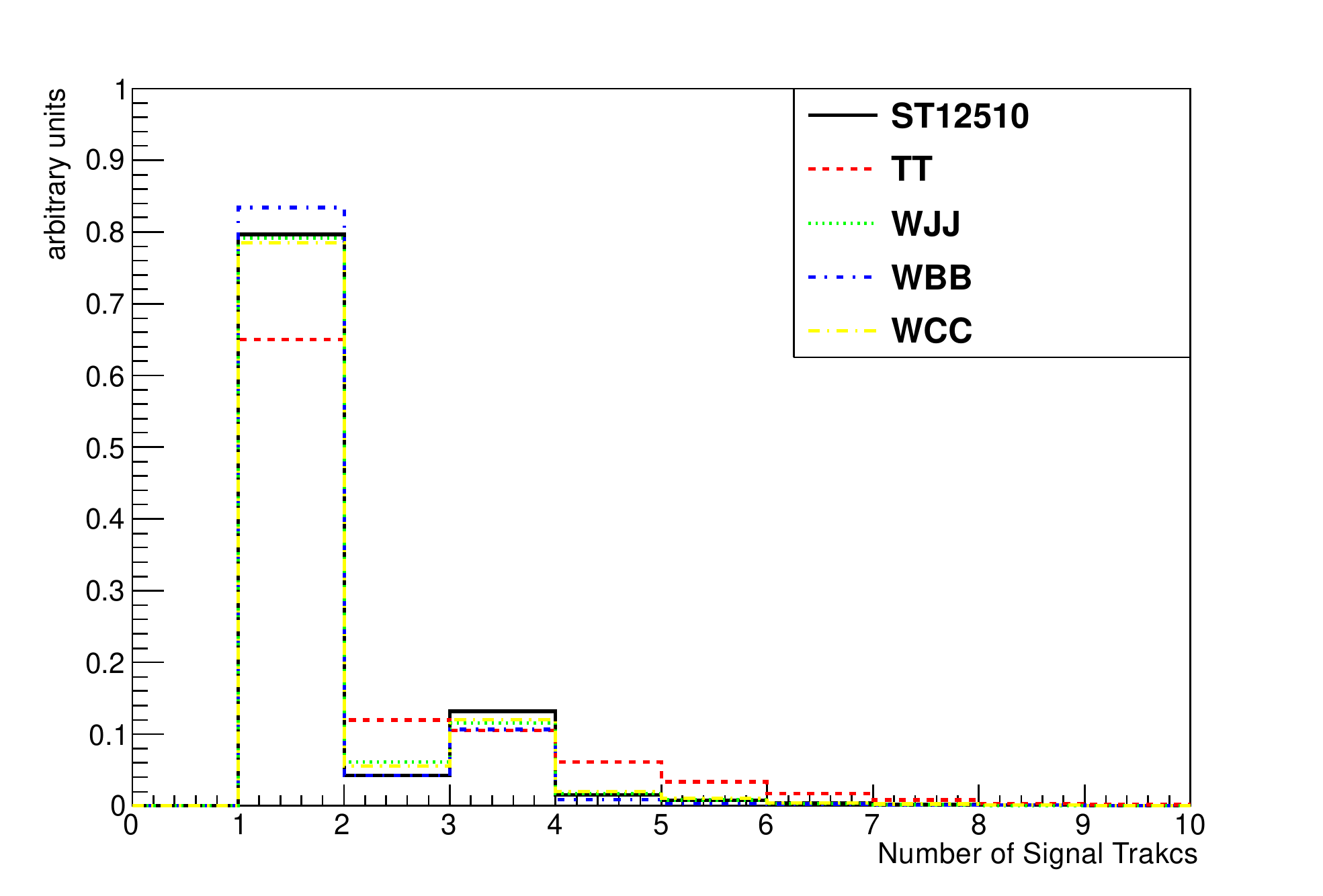}
\end{center}
\caption{Number of tracks in the signal cone of the $\tau$-jet cnadidate in signal and background events.}
\label{nstk}
\end{figure}

\begin{figure}
\begin{center}
\includegraphics[width=0.6\textwidth]{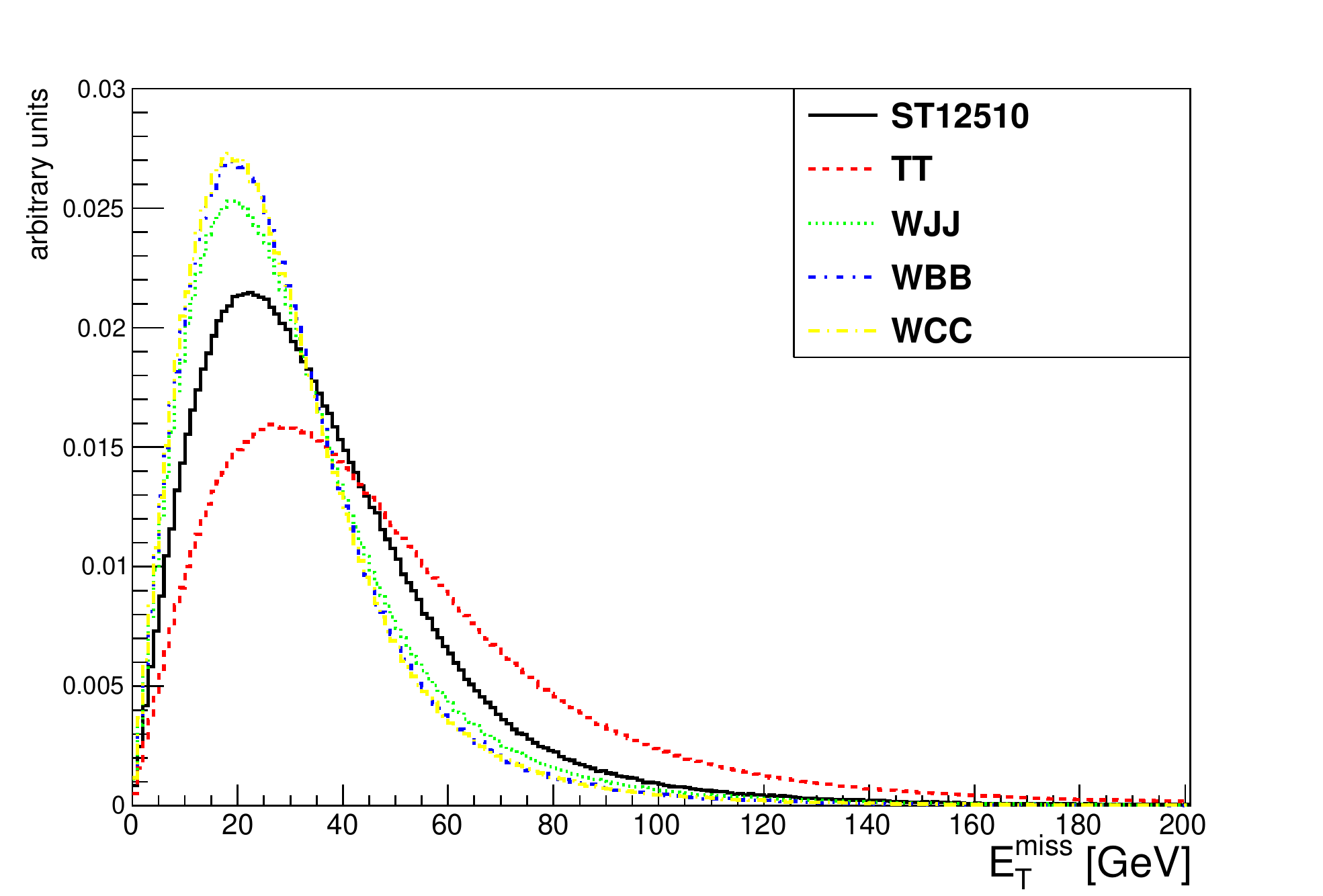}
\end{center}
\caption{Missing $E_{T}$ distribution in signal and background events.}
\label{met}
\end{figure}

\begin{figure}
\begin{center}
\includegraphics[width=0.6\textwidth]{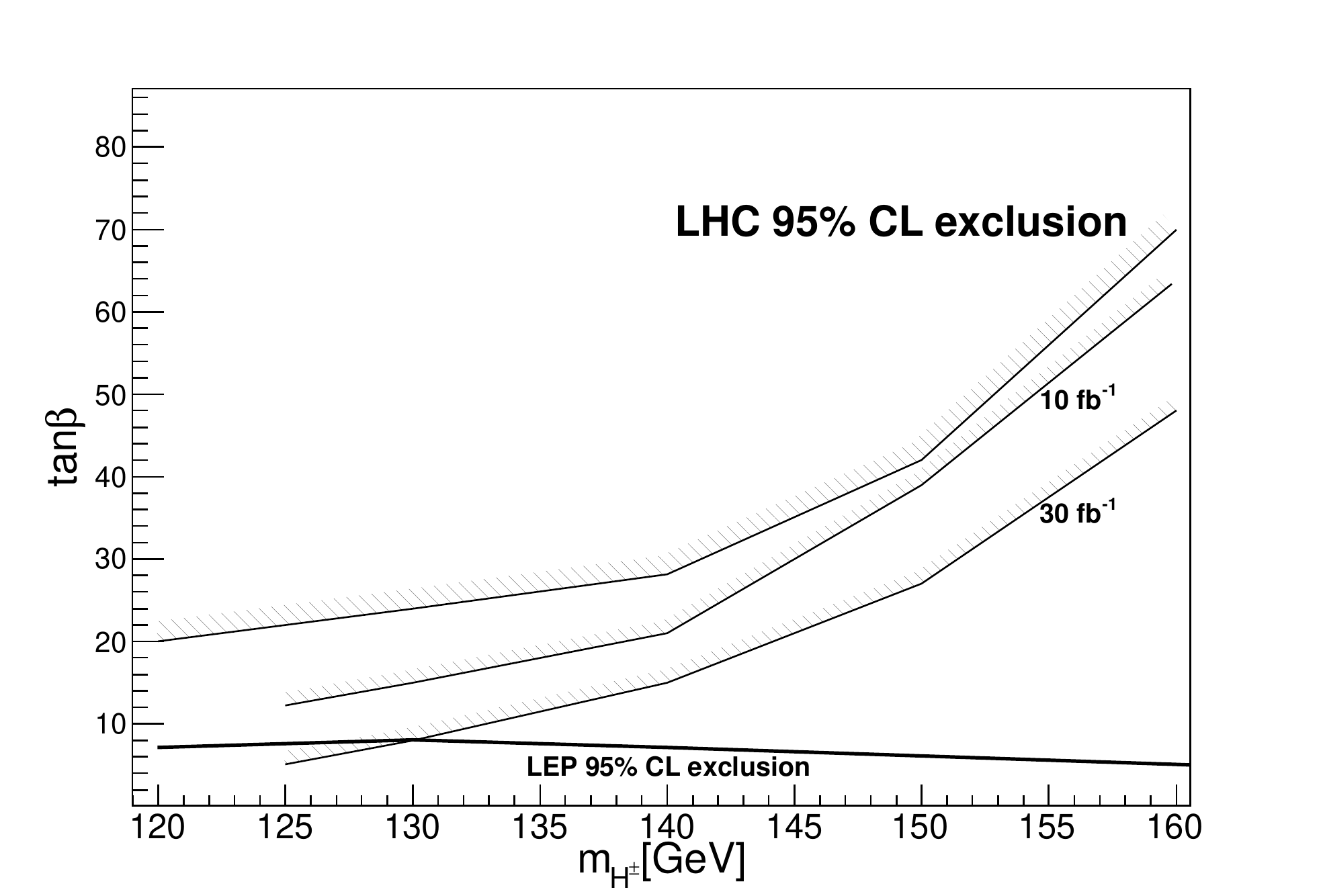}
\end{center}
\caption{Excluded area at 95$\%$ C.L. for different integrated luminosities.}
\label{limit}
\end{figure}

\begin{figure}
\begin{center}
\includegraphics[width=0.6\textwidth]{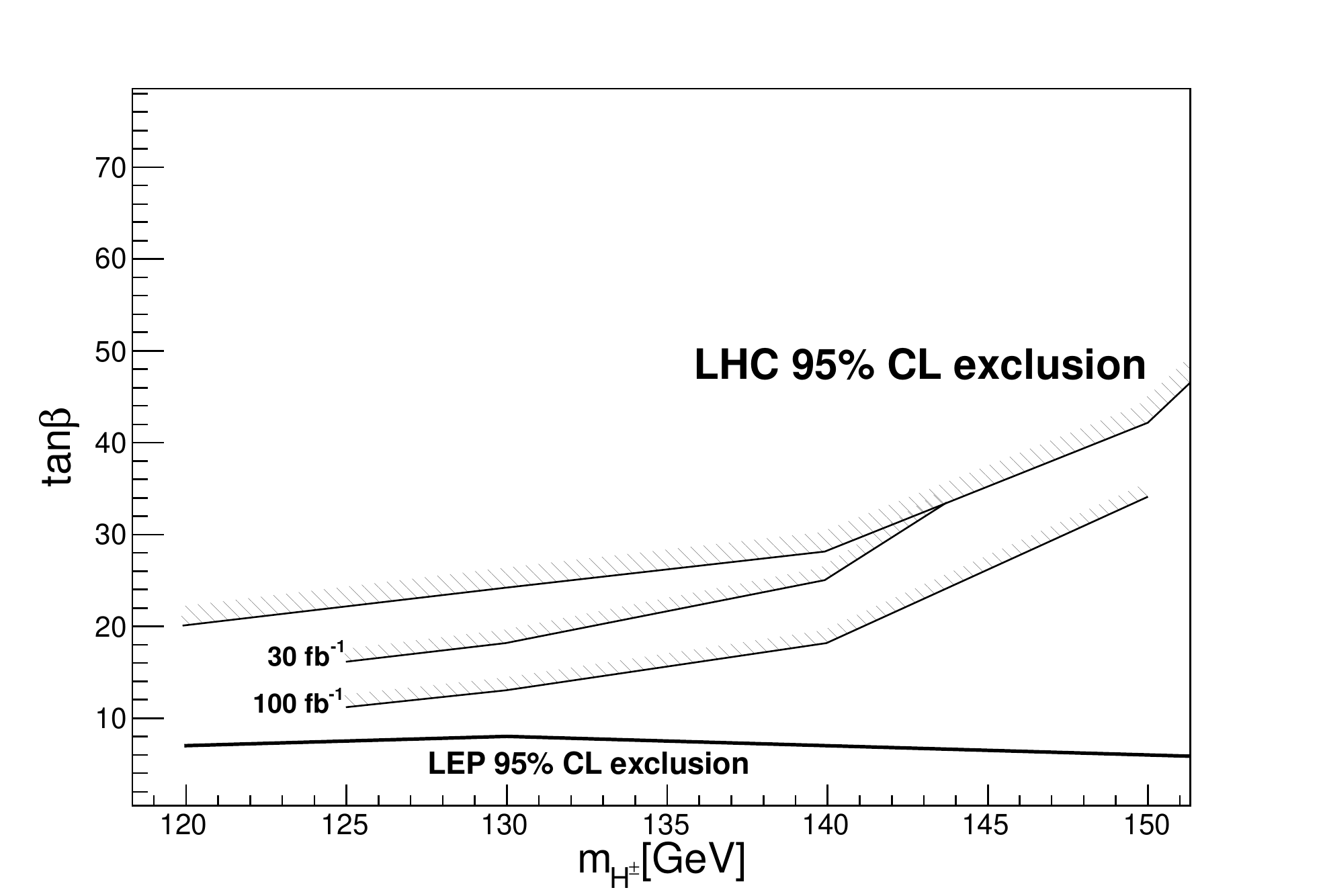}
\end{center}
\caption{The $5\sigma$ contour for different integrated luminosities.}
\label{5s}
\end{figure}

Having generated events, a hadron level analysis is designed based on kinematic and topological differences of signal and background events. The main features of signal events are low jet multiplicity (three jets including the $\tau$-jet), and pseudorapidity distribution of the light jet.\\
The low jet multiplicity is a feature of signal events which can be used to suppress $t\bar{t}$ events and single W events accompanied by more than two jets. Figure \ref{jm} compares signal and background events in terms of their jet multiplicities. Throughout the paper, plots are shown with a signal comprising of a charged Higgs with mass $m_{H^{\pm}}=125$ GeV and \tanb = 10. The abbreviation used for the signal is ``ST12510''.
In Fig. \ref{jm}, jets with $E_{T}>20$ GeV and $|\eta|<5$ are included. Events are selected for further processing if the number of all jets equals three.\\
The b-jets are reconstructed with a matching algorithm in which a jet is tagged as a b-jet if $\Delta R$ between the jet and a $b$ or $c$ quark with $p_{T}>20$ GeV and $|\eta|<2.5$ is less than 0.4. The $b$-jet efficiency is assumed to be 40 $\%$ while the $c$-jet mistagging rate is taken to be 10$\%$.\\
Figure \ref{bjm} shows $b$-jet multiplicity in signal and background events.
The shown distribution has a little dependence on the charged Higgs mass in signal events. With a heavier charged Higgs, a smaller phase space is available to the $b$-quark in the top quark decay, thus suppressing the $b$-jet kinematically.\\
The $t$-channel single top involves a light quark which proceeds with almost small deviation during the W exchange. Therefore signal events are characterized with a forward jet close to the beam pipe with a pseudorapidity distribution which has maxima in the high $|\eta|$ region of the detector. In order to select these light jets, jets close to light quarks within $\Delta R<0.4$ are selected. Figure \ref{ljet} and \ref{ljeta} shows their $E_{T}$ and $|\eta|$ distributions. The difference in $|\eta|$ distributions between the signal and background events is now clear. This feature is used to suppress background events by applying the kinematic requirement as in Eq. \ref{ljetaeq}. The number of light jets which satisfy Eq. \ref{ljetaeq} is counted and distributions of light jets is plotted as in Fig. \ref{ljetmul}. There must be one and only one light jet in the event which satisfies Eq. \ref{ljetaeq}.  
\be
E_{T}>20~\textnormal{GeV},~|\eta|_{\textnormal{light jet}}>3
\label{ljetaeq}
\ee
In addition to the cut applied on the light jet pseudorapidity distribution, angular correlations can also be used to further increase the signal to background ratio. Such studies have been performed on single top events in the leptonic final state in \cite{atlasst} and \cite{sttevatron}. One of such angular variables, i.e., $\Delta \eta$ between the $b-$jet and the $\tau-$jet was studied in this analysis as shown in Fig. \ref{detabt}. Applying a cut on this distribution as $|\Delta \eta|<1.5$ results in the following efficiencies for signal ($m_{H^{\pm}}=125$ GeV), $t\bar{t}$, W+2jets, $Wbb$ and $Wcc$ respectively: 0.84, 0.63, 0.61, 0.71, 0.76. These numbers indicate that S/B increases with the applied cut, however, the signal significance may not increase sizably. Including only W+2jets as the main background, a significance evaluated as $S/\sqrt{B}$ increases by a factor 1.07, i.e., an increase of the order of $7\%$ or less is expected when including angular variables in this analysis. This gain was not included in the final conclusions as it does not change the final contours sizably.\\
The next step is to select $\tau$-jets through their hadronic decays. The $\tau$ identification algorithm used in this analysis follows the standard algorithm developed at LHC \cite{tauid}. Since $\tau$ jets produce low charged particle multiplicity in their hadronic decays, the leading track (the track with highest $p_{T}$) in the $\tau$-jet cone is expected to be hadrder than the corresponding tracks in light jets. Therefore the following requirement on the leading track is applied.
\be
\textnormal{Leading track} ~p_{T}>5~ \textnormal{GeV}
\label{ltk}
\ee 
The isolation requirement applied on $\tau$-jets is based on the fact that the $\tau$-jet accomodates few charged tracks close to the central axis of the jet cone. Therefore no charged tracks with $p_{T}$ above a threshold (which is taken to be 1 GeV) should be in the annulus defined as in Eq. \ref{isol}.
\be
\textnormal{No tracks with} ~p_{T}>1~ \textnormal{GeV in}~ 0.07<\Delta R<0.4 
\label{isol}
\ee
The annulus is defined around the leading track in the $\tau$-jet cone. Since $\tau$-jets predominantly decay to one or three charged pions, one would expect one or three charged tracks in the signal cone defined as $\Delta R < 0.07$ around the leading track. The distribution of signal cone tracks for signal and background events is shown in Fig. \ref{nstk}. 
The requirement on the number of signal cone tracks is thus defined as in Eq. \ref{nstkeq}.
\be
\textnormal{Number of signal cone tracks = 1 or 3} 
\label{nstkeq}
\ee
There should be one and only one jet passing all the above requirements of the $\tau$-identification algorithm. As the last step the missing transverse energy is calculated as the negative sum over particle momenta in $x$ and $y$ directions. Figure \ref{met} shows the missing $E_{T}$ distributions in signal and background events. A cut is applied on missing $E_{T}$ as in Eq. \ref{meteq}. 
\be
\textnormal{Missing}~E_{T}> 20 ~\textnormal{GeV}
\label{meteq}
\ee
Application of a harder cut may also be studied at the expense of decreasing the signal statistics. Now signal and background events pass through all above requirements and selection efficiencies are obtained as in Tab. \ref{seleffs}. The signal events in Tab. \ref{seleffs} correspond to three charged Higgs mass hypotheses $m_{H^{\pm}}=125,~140,~160$ GeV and \tanb = 10. As mentioned before, the $s$-channel single top has no sizable contribution. Taking $m_{H^{\pm}}=125$ GeV and \tanb = 10 as an example, $\sigma(s\textnormal{-channel})=10.5~pb$, BR$(t\ra H^{+}b)=0.007$ and a total selection efficiency of 0.004 is obtained which leads to 0.9 events at 30 $fb^{-1}$. The $t\bar{t}$ process may also be regarded as a source of charged Higgs bosons. When one of the top quarks decays to charged Higgs, i.e., $t\bar{t}\ra H^{\pm}W^{\mp}b\bar{b}\ra \tau\nu j j b\bar{b}$, the final state consists of five jets. Such an event is unlikely to be selected by the single top selection described in this paper. It was explicitly checked and turned out that these events pass the kinematic selection cuts with selection efficiency of $7\times 10^{-5}$. With $m_{H^{\pm}}=125$ GeV, \tanb = 10 and BR$(t\ra H^{+}b)=0.007$, a total number of 16 events of this type would be expected at 30$fb^{-1}$. This amount, though being part of the signal, is negligible compared to what is obtained from single top events (364 events at the same integrated luminosity). Based on efficiencies listed in Tab. \ref{seleffs}, exclusion limits are obtained at the 95$\%$ C.L. using TLimit class implemented in ROOT \cite{root}. Figure \ref{limit} shows the 95 $\%$ C.L. exclusion area for integrated luminosities of 10 and 30 $fb^{-1}$ compared to the current LHC exclusion obtained from direct searches. As Fig. \ref{5s} shows, the $5\sigma$ contour covers areas not excluded by direct searches, thus providing the opportunity for observation of the signal at $5\sigma$ statistical significance. 

\begin{table}[h]
\begin{center}
\begin{tabular}{|c|c|c|c|c|c|c|c|}
\hline
Samples & ST12510 & ST14010 & ST16010 & $t\bar{t}$ & $Wjj$ & $Wb\bar{b}$ & $Wc\bar{c}$ \\
$\sigma\times \textnormal{BR} ~[pb]$ & 1.73 & 1.16 & 0.18 & 122.5 & 1670 & 3.5 & 3.3 \\
3 jets & 0.25 & 0.23 & 0.22 & 0.21 & 0.22 & 0.2 & 0.21 \\
1 b-jet & 0.44 & 0.32 & 0.21 & 0.13 & 0.11 & 0.47 & 0.45 \\
1 light jet & 0.22 & 0.22 & 0.21 & 0.043 & 0.069 & 0.015 & 0.054 \\
Leading track (Eq. \ref{ltk}) & 0.98 & 0.98 & 0.99 & 1 & 0.99 & 0.97 & 0.99 \\
Isolation (Eq. \ref{isol})& 0.42 & 0.44 & 0.43 & 0.14 & 0.31 & 0.3 & 0.35 \\
Signal tracks (Eq. \ref{nstkeq})& 0.93 & 0.93 & 0.92 & 0.58 & 0.88 & 0.89 & 0.89 \\
1 tau & 0.97 & 0.98 & 0.98 & 0.98 & 0.96 & 0.97 & 0.97 \\
Met (Eq. \ref{meteq}) & 0.78 & 0.82 & 0.87 & 0.82 & 0.71 & 0.71 & 0.71 \\
Total eff & 0.007 & 0.005 & 0.003 & 7.7 $\times 10^{-5}$ & 3 $\times 10^{-4}$ & 2.6$\times 10^{-4}$ & 0.001  \\
Events at 30$fb^{-1}$ & 364 & 174 & 16.2 & 283 & 15030 & 27.3 & 99\\
\hline
\end{tabular}
\end{center}
\caption{Selection efficiencies. Labels for the signal events (ST12510, ST14010 and ST16010) correspond to charged Higgs masses $m_{H^{\pm}}=125,~140,~160$ GeV and \tanb = 10. \label{seleffs}}
\end{table}
\section{Discussion on the Background Control}
As Figs. \ref{limit} and \ref{5s} show, promising results are obtained in terms of exclusion or discovery at a reasonable amount of integrated luminosity. However, according to Tab. \ref{seleffs}, the signal to background ratio (S/B) is small mainly due to the large remaining W+2jets background. Therefore a reasonable understanding of this background is required for reliability of the results.\\ 
This background can in principle be suppressed by requiring two b-tagged jets in the event, as the incoming b-jet is usually originated from a gluon splitting which produces two b-jets. One of those b-jets escapes detection or is detected in forward directions at a high $\eta$. Requiring two b-jets in the event increases S/B, but may suppress the signal as well. In full analyses, both strategies (one and two b-tagged jets) are used and results are compared. Since the studied channel in this analysis has a small cross section due to the small branching ratio of top decay to charged Higgs, we did not try two b-tagging requirement. However, one can repeat the analysis using this requirement and evaluate how the significance and S/B change with this requirement.\\
The amount of W+2jets background can also be controlled by defining control regions which enhance this background compared to other processes. The usual procedure is to remove the b-tagging requirement and study the W+2jets background shape in a distribution such as the W boson invariant mass. Such a distribution is expected to be dominated by W bosons from W+2jets events and the shape of this distribution can be used to estimate the W+2jets background using real data and measure its uncertainty as done in \cite{d0st}. Any conclusion about the signal observability cretainly depends on how well we can estimate the W+2jets background from data, however, a detailed study beyond the scope of this analysis is needed to achieve this goal.\\
Another point to mention is that although the W+2jets background has been the main concern in single top searches in all experiments, it has been under control in cut-based or multivariate analyses \cite{d0st,cdfst,atlasst}. The high amount of fake rate from W+2jets in this analysis is due to the very basic $\tau$-identification algorithm which leaves a high fake rate from light quark jets and also the simple matching-based $b-$tagging algorithm used in the analysis. It is expected that at a full analysis with state-of-the-art algorithms developed for $\tau$-jets and $b-$jets, the fake rate is dramatically suppressed and a smaller number of W+2jets events remain. Therefore a full detector analysis is motivated and needed to confirm the results obtained in this work.               
\section{Conclusions}
The single top production was analyzed as a source of charged Higgs bosons. It was shown that at $\sqrt{s}=14$ TeV, corresponding to LHC nominal center of mass energy, single top events have enough cross sections and unique kinematic features to be used for exclusion or observation of light charged Higgs. The low jet multiplicity and forward light jets are among the kinematic features of signal events which helps distinguishing the signal from the background events. These tools were used in the analysis and it was shown that what can be obtained in terms of exclusion at 95$\%$ C.L. or 5$\sigma$ discovery proceeds LHC results. Finally 95$\%$ C.L. exclusion and 5$\sigma$ contours were plotted in $m_{H^{\pm}},tan\beta$ plane.

\end{document}